\documentclass[preprint,prc,showpacs,preprintnumbers,amsmath,amssymb,floatfix]{revtex4}
\usepackage{hyperref}
\usepackage{booktabs}
\usepackage{threeparttable}
\usepackage{amsmath}
\usepackage{graphicx}
\usepackage{dcolumn}
\usepackage{bm}
\usepackage{ulem} 
\usepackage[usenames]{color}
\usepackage{epstopdf}
\usepackage{epsfig}
\usepackage{float}
\usepackage{subfigure}
\usepackage{multirow}
\usepackage[version=3]{mhchem}

\newcommand{\nc}{\newcommand}       
\nc{\vc}[1] {\mbox{\boldmath $#1$}} 
\nc{\del}       {\partial}              
\nc{\bra}       {\langle}               
\nc{\ket}       {\rangle}               
\nc{\bras}[1]   {\langle #1|}           
\nc{\kets}[1]   {|#1\rangle}            
\nc{\mapleft}[1]{           
 \smash{\mathop{\,          %
  \hbox to 1.5cm{\rightarrowfill}\, }\limits_{#1}}}
\nc{\beq}     {\begin{eqnarray}}
\nc{\eeq}    {\end{eqnarray}}
\nc{\nn}      {\\\nonumber}
\nc{\vs}      {\vspace{-0.275cm}}
\nc{\fra}    {\frac{1}{2}}
\nc{\mb}        {\mathbf}
\begin{document}

\preprint{}

\title{Single $\Lambda_c^+$ hypernuclei within quark mean-field model}
\author{Linzhuo Wu}
\affiliation{School of Physics, Nankai University, Tianjin 300071,  China}
\author{Jinniu Hu}
\email{hujinniu@nankai.edu.cn}
\affiliation{School of Physics, Nankai University, Tianjin 300071,  China}
\author{Hong Shen}
\affiliation{School of Physics, Nankai University, Tianjin 300071,  China}

\date{\today}
\begin{abstract}
The quark mean-field (QMF) model is applied to study the single $\Lambda^+_c$ hypernuclei. The charm baryon, $\Lambda^+_c$, is constructed by three constituent quarks, $u, ~d$, and $c$, confined by central harmonic oscillator potentials. The confinement potential strength of charm quark is determined by fitting the experimental masses of charm baryons, $\Lambda^+_c,~\Sigma^+_c$, and $\Xi^{++}_{cc}$. The effects of pions and gluons are also considered to describe the baryons at the quark level. The baryons in $\Lambda^+_c$ hypernuclei interact with each other through exchanging the $\sigma,~\omega$, and $\rho$ mesons between the quarks confined in different baryons. The $\Lambda^+_c N$ potential in the QMF model is strongly dependent on the coupling constant between $\omega$ meson and $\Lambda^+_c$, $g_{\omega\Lambda^+_c}$. When the conventional quark counting rule is used, i. e., $g_{\omega\Lambda^+_c}=2/3g_{\omega N}$, the massive $\Lambda^+_c$ hypernucleus can exist, whose single $\Lambda^+_c$ binding energy is smaller with the mass number increasing due to the strong Coulomb repulsion between $\Lambda^+_c$  and protons. When $g_{\omega\Lambda^+_c}$ is fixed by the latest lattice $\Lambda^+_c N$ potential, the $\Lambda^+_c$ hypernuclei only can exist up to $A\sim 50$.
\end{abstract}



\maketitle

\section{Introduction}

The strangeness degree of freedom was studied from the early 1950s to explain the strange particles and hypernucleus observed in the cosmic rays~\cite{danysz53}. After the quark models were proposed by Gell-Mann and Zweig in 1960s, it was regarded that the strangeness in nuclear physics was generated by the strange ($s$) quark. With the developments of accelerators and detectors, many $\Lambda$ hypernuclei with $\Lambda$ hyperon bound in nuclei were observed in the large nuclear facilities in the past half century~\cite{hashimoto06,feliciello15,gal16} {from $_{\Lambda}^{3}\rm {H}$ to $_{\Lambda}^{208}\rm {Pb}$. The $\Sigma$ hypernuclei were not detected except the $_{\Sigma}^{4}\rm {He}$ quasibound state~\cite{hayano89,nagae98}. It was generally considered that the $\Sigma N$ interaction is repulsive. Furthermore}, there were also some experimental evidences to indicate the existence of $\Xi$ hypernuclei~\cite{aoki93,khaustov00,yamaguchi01,nakazawa15,gogami16} {and few $\Lambda\Lambda$ light hypernuclei~\cite{danysz63,prowse66,aoki91,ahn13}}. 

The hyperons do not have to obey the Pauli exclusion principle in normal nuclear system, which can be easily bound in a nucleus. Therefore, the hypernucleus is a good probe to investigate the baryon-baryon interaction~\cite{vidana00,vidana01,hiyama09,hiyama10}. Many new-generation facilities, such as, FAIR, JLab, J-PARC, MAMI, and HIAF are planning to explore more unknown $\Lambda$ hypernuclei in the nuclear landscape~\cite{feliciello15}. In the aspect of theoretical researches, various nuclear models were applied to study the hypernuclei, such as the {\it ab initio} methods for light hypernuclei~\cite{hiyama08,wirth14}, {$G$-matrix calculation~\cite{vidana04}}, shell model~\cite{millener08}, Skyrme-Hartree-Fock model~\cite{li13,schulze13,cui15,zhou16}, relativistic mean-field model~\cite{yuichi94,mares94,shen06,xu12,sun16,fortin17,liu18}, quark meson-coupling model~\cite{tsushima97,tsushima98,saito07}, quark mean-field model~\cite{shen02,hu14a,hu14b}, and so on for heavy hypernuclei. These models can describe the ground-state properties of $\Lambda$ hypernuclei very well with various effective $\Lambda N $ interactions.

In addition to the up, down, and strange quarks, there are also charm, bottom, and top quarks in the universe, which can combine with the up and down quarks to constitute exotic baryons. The $\Lambda^+_c$ was the first charmed baryon confirmed in experiment, whose components are very similar to the $\Lambda$ hyperon~\cite{patrignani16}.  Only the strange quark is replaced by charm quark in  $\Lambda^+_c$. A natural question is whether  $\Lambda^+_c$ and normal nuclei can bind together to form a charmed hypernuclei. Actually, forty years ago, Dover and Kahana already discussed the possibility of charmed hypernuclei with a $\Lambda^+_c N$ potential generated by $SU(4)$ symmetry, where the bound states of a charmed baryon and normal nuclei were predicted~\cite{dover77}. Then, the light charmed hypernuclei were investigated by {cluster model and few-body methods}~\cite{bhamathi81,bando82,gibson83}. The heavy nuclei are better described by the density functional theory. Accordingly, the massive charmed hypernuclei were calculated by the quark meson-coupling (QMC) model~\cite{tushima03a,tushima03b,tushima04,tushima18} and relativistic mean-field (RMF) model~\cite{tan04,tan04a}. The binding energies, density distribution, impurity effect, medium effect of charmed hypernuclei were widely discussed in these works. {Meanwhile, the investigations of $\Lambda^+_c$ hypernuclei in aspect of experiment were explored in 1970s and 1980s in Dubna, which only reported three possible candidate events due to the difficult production mechanism of charmed hypernuclei~\cite{batusov76,batusov81,lyukov89}. In future, FAIR and JPARC are hopefully expected to produce sufficient charmed particles to generate more charmed hypernuclei~\cite{riedl07,shyam17,krein18}.}

The essential element to determine the properties of charmed hypernuclei is the strength of $\Lambda^+_c N$ potential. In the early time, it was obtained by extending the one-boson-exchange potential (OBEP) for nucleon-nucleon and nucleon-hyperon systems with $SU(4)$ symmetry~\cite{dover77}. Recently, Liu and Oka considered a more reasonable Lagrangian of OBEP to include the chiral symmetry, heavy quark symmetry, and hidden local symmetry~\cite{liu12,maeda18}. In QMC and RMF models, the coupling constants between charm baryons and mesons were usually generated by the naive quark counting rules.

The more reliable and cheerful progress about $\Lambda^+_c N$ potential was from the lattice QCD simulation. HAL QCD Collaboration calculated the central and tensor components of $\Lambda^+_c N$ potential at $^1S_0$ and $^3S_1$-$^3D_1$ channels within $(2+1)$-flavor lattice QCD at quark masses corresponding to pion masses, $m_\pi\simeq 410,~570,~700$ MeV, respectively. It was found that the $\Lambda^+_c N$ potentials with such quark masses were attractive at $^1S_0$ and $^3S_1$ channels~\cite{inoue11,sasaki15,miyamoto16,miyamoto17,miyamoto18}. Later, Haidenbauer and Krein extrapolated the $\Lambda^+_c N$ potential at physical pion mass with chiral effective field theory from the HAL QCD results at large quark masses. They also claimed that the $\Lambda^+_c N$ potential at $m_\pi=138$ MeV could make the four-body and five-body charmed hypernuclei bind~\cite{haidenbauer18}. With these achievements, Miyamoto {\it et al.} derived a single-folding $\Lambda^+_c N$ potential for $\Lambda^+_c$ hypernuclei generated by lattice QCD simulation, where the $\Lambda^+_c$ hypernuclei could exist between the mass numbers from $A=12$ to $A\sim50$~\cite{miyamoto18}. Furthermore, Vida\~{n}a {\it et al.} recently also discussed the charmed hypernuclei within a microscopic many-body approach with an $SU(4)$ extension of OBEP {from the J\"ulich hyperon-nucleon potential~\cite{vidana19}. It was found that the phase shifts from the B and C models of Ref.\cite{vidana19} agree to those extracted from HAL QCD data at physical pion mass by Haidenbauer and Krein. Furthermore, their results about charmed hypernuclei were also compatible with other theoretical calculations~\cite{tushima03a,tushima03b,tushima04,tushima18,tan04,tan04a}}.

The quark mean-field (QMF) model is a very powerful nuclear many-body method from the quark level. The baryon is regarded to be constructed by three constituent quarks with central confinement potentials. The baryon-baryon interaction in nucleus is realized by exchanging the $\sigma,~\omega$, and $\rho$ mesons between the quarks in different baryons. The QMF model has been successfully used to study the properties of normal nuclei, $\Lambda,~\Xi$ hypernuclei and neutron star after including the effects of pions and gluons at hadron level~\cite{barik13,mishra15,mishra16,xing16,xing17,hu17}.

In this work, we would like to apply the QMF model to study the properties of charmed hypernuclei, especially $\Lambda^+_c$ ones. The $\Lambda^+_c$ baryon consists of $u,~d$, and $c$ quarks, which are confined by the central harmonic oscillator potentials. The strength of confinement potential for charm quark will be fixed by the experimental masses of charmed baryons. The coupling constants between charm quark and mesons will be determined by two schemes. The first one is decided by the naive quark counting rules. The second one is extracted from the HAL QCD simulations.  This article is organized as follows. In section II, the theoretic framework of QMF model related to charmed hypernuclei is presented. The results and discussions for $\Lambda^+_c$ hypernuclei will be shown in section III. The summary and conclusions will be given in section IV.

\section{Quark mean-field model for charmed hypernuclei}

In this section, we will give a brief introduction of QMF model for charmed hypernuclei. In QMF model, baryons are composed of three constituent quarks, which are confined by the central confinement potentials. The specific form of such potentials cannot be
obtained directly because of the non-perturbative character of QCD theory in low-energy region. Many phenomenological confinement potentials have been proposed, where the polynomial forms were widely used. In this work, we adopt a harmonic oscillator potential with a mixing scalar-vector structure~\cite{mishra15,mishra16,xing16,xing17},
\beq\label{1}
U_q(r)=\frac{1}{2}(1+\gamma^0)(a_qr^2+V_q),
\eeq
where the potential parameters $a_q$ and $V_q$ will be determined by the masses of charmed baryons and $q$ denotes $u$, $d$ or $c$, respectively. In this case, the Dirac equation including the nuclear medium effect for confined quark is written as
\beq\label{2}
[\gamma^{0}(\epsilon_{q}-g_{\omega q}\omega-\tau_{3}g_{\rho q}\rho)-\vec{\gamma}\cdot\vec{p}-(m_{q}
-g_{\sigma q}\sigma)-U_q(r)]\psi_{q}(\vec{r})=0.
\eeq
Here, $\psi_{q}(\vec{r})$ represents the quark field. $\sigma, ~\omega$, and $\rho$ are the classical meson fields, which are exchanged between quarks in different baryons to achieve the baryon-baryon interaction. $g_{\sigma q}, ~g_{\omega q}$, and $g_{\rho q}$ are the coupling strengths of $\sigma,~\omega$, and $\rho$ mesons with quarks, respectively. $m_q$ is the constituent quark mass and $\tau_{3}$ corresponds to the third component of isospin matrix. This equation can be solved exactly and its ground-state solution of the energy satisfies the eigenvalue condition
\beq\label{3}
(\epsilon_q'-m_q')\sqrt{\frac{\lambda_q}{a_q}}=3,
\eeq
where
\beq\label{4}
\epsilon_q'=&&\epsilon_q^\ast-V_q/2,\nn
m_q'=&&m_q^\ast+V_q/2,\nn
\lambda_q=&&\epsilon_q'+m_q'=\epsilon_q^\ast+m_q^\ast.
\eeq
Considering the effect of nuclear medium generated by the meson fields, the effective single-quark energy and effective quark mass are defined by
\beq\label{5}
&&\epsilon_q^*=\epsilon_{q}-g_{\omega q}\omega-\tau_{3}g_{\rho q}\rho,\nn
&&m_q^*=m_{q}-g_{\sigma q}\sigma.
\eeq
 The corresponding wave function is
 \beq\label{6}
 \psi_q=\frac{1}{\sqrt{4\pi}}
         \left(
               \begin{array}{c}
                ig_q(r)/r\\
                \vec{\sigma}\cdot\hat{\vec{r}}f_q(r)/r\\
               \end{array}
             \right)\chi_{s},
\eeq
where
\beq\label{7}
g_q(r)=&&\mathcal{N}_q(\frac{r}{r_{0q}})e^{-r^2/2r_{0q}^2},\nn
f_q(r)=&&-\frac{\mathcal{N}_q}{\lambda_qr_{0q}}(\frac{r}{r_{0q}})^2e^{-r^2/2r_{0q}^2}.
\eeq
The normalization constant has $\mathcal{N}_q^2=\frac{8\lambda_q}{\sqrt{\pi}r_{0q}}\frac{1}{3\epsilon_q'+m_q'}$ and $r_{0q}=(a_q\lambda_q)^{-1/4}$.
The ground-state energy for quark $\epsilon_q^\ast$ can be obtained by solving Eq.(\ref{3}). Accordingly, the binding energy of three quarks as the zeroth-order energy of the baryon can be written immediately as,
\beq\label{8}
E_B^{*0}=\sum_q\epsilon^*_q.
\eeq

Three corrections should be taken into account based on the zeroth-order energy of the baryon, including the center-of-mass correction $\epsilon_{\mathrm{c.m.}}$, the pion correction $\delta M^\pi_B$ and the gluon correction $\left(\Delta E_{B}\right)_{g}$ to generate the real baryon mass. The center-of-mass correction should be considered due to the translation invariance of baryons. The pion correction comes from the restoration of chiral symmetry of QCD theory. The gluon correction is generated by the short-range exchanging interaction among quarks. These three corrections are formulated in detail as following~\cite{mishra15,xing16,xing17}.

The energy contribution of center-of-mass correction can be written as
\beq\label{9}
\epsilon_{\rm c.m}=e_{\rm c.m}^{(1)}+e_{\rm c.m}^{(2)},
\eeq
where
\beq\label{10}
e_{\mathrm{c.m.}}^{(1)}=&&\sum^3_{i=1}\left[\frac{m_{q_{i}}}{\sum^3_{k=1}m_{q_{k}}}\frac{6}{r^2_{0q_{i}}(3\epsilon'_{q_{i}}+m'_{q_{i}})}\right],\nn
e_{\mathrm{c.m.}}^{(2)}=&&\frac{1}{2}\left[\frac{2}{\sum_{k} m_{q_{k}}} \sum_{i} a_{i} m_{i}\left\langle r_{i}^{2}\right\rangle+\frac{2}{\sum_{k} m_{q_{k}}}\sum_{i}a_{i} m_{i}\left\langle \gamma^{0}(i) r_{i}^{2}\right\rangle \right. \nn
&&\left.-\frac{3}{\left(\sum_{k}m_{q_{k}}\right)^{2}} \sum_{i} a_{i} m_{i}^{2} \left\langle r_{i}^{2}\right\rangle-\frac{1}{\left(\sum_{k} m_{q_{k}}\right)^{2}}\sum_{i}\left\langle\gamma^{0}(1) a_{i} m_{i}^{2} r_{i}^{2}\right\rangle \right. \nn
&&\left.-\frac{1}{\left(\sum_{k} m_{q_{k}}\right)^{2}} \sum_{i}\left\langle\gamma^{0}(2) a_{i} m_{i}^{2} r_{i}^{2}\right\rangle-\frac{1}{\left(\sum_{k} m_{q_{k}}\right)^{2}} \sum_{i}\left\langle\gamma^{0}(3) a_{i} m_{i}^{2} r_{i}^{2}\right\rangle \right].
\eeq
The expectation values associated with the radii are evaluated as following,
\beq\label{11}
\left\langle r_{i}^{2}\right\rangle=&&\frac{\left(11 \epsilon_{q i}^{\prime}+m_{q i}^{\prime}\right) r_{0 q i}^{2}}{2\left(3 \epsilon_{q i}^{\prime}+m_{q i}^{\prime}\right)}, \nn
\left\langle\gamma^{0}(i) r_{i}^{2}\right\rangle=&&\frac{\left(\epsilon_{q i}^{\prime}+11 m_{q i}^{\prime}\right) r_{0 q i}^{2}}{2\left(3 \epsilon_{q i}^{\prime}+m_{q i}^{\prime}\right)}, \nn
\left\langle\gamma^{0}(i) r_{j}^{2}\right\rangle_{i \neq j}=&&\frac{\left(\epsilon_{q i}^{\prime}+3 m_{q i}^{\prime}\right)\left\langle r_{j}^{2}\right\rangle}{3 \epsilon_{q i}^{\prime}+m_{q i}^{\prime}}.
\eeq

The energy contributions of pion correction for nucleon and charmed baryons $\Lambda_c^+,~\Sigma_c^+,~\Xi_{cc}^{++}$ are given by
\beq\label{12}
\delta M_{N}^{\pi}=&&-\frac{171}{25} f_{N N \pi}^{2} I_{\pi} ,\nn
\delta M_{\Lambda_c^+}^{\pi}=&&-\frac{108}{25} f_{N N \pi}^{2} I_{\pi},\nn
\delta M_{\Sigma_c^+}^{\pi}=&&-\frac{12}{5} f_{N N \pi}^{2} I_{\pi},\nn
\delta M_{\Xi_{cc}^{++}}^{\pi}=&&-\frac{27}{25} f_{N N \pi}^{2} I_{\pi},
\eeq
where
\beq\label{13}
I_{\pi}=\frac{1}{\pi m_{\pi}^{2}} \int_{0}^{\infty} d k \frac{k^{4} u^{2}(k)}{w_{k}^{2}},
\eeq
and the axial vector nucleon form factor is written as
\beq\label{14}
u(k)=\left[1-\frac{3}{2} \frac{k^{2}}{\lambda_{u}\left(5 \epsilon_{u}^{\prime}+7 m_{u}^{\prime}\right)}\right] e^{-\frac{1}{4} r_{0 u}^{2} k^{2}}.
\eeq
The pseudovector $N\pi$ coupling constant $f_{NN\pi}$ can be derived from the Goldberg-Triemann relation
\beq\label{15}
f_{N N \pi}=\frac{25 \epsilon_{u}^{\prime}+35 m_{u}^{\prime}}{27 \epsilon_{u}^{\prime}+9 m_{u}^{\prime}} \frac{m_{\pi}}{4 \sqrt{\pi} f_{\pi}},
\eeq
where $m_{\pi}=140$ MeV and $f_{\pi}=93$ MeV are the pion mass and the phenomenological pion decay constant, respectively.

The energy contribution from gluon correction in baryon mass consists of a color electric part and a magnetic part as
\beq\label{16}
\left(\Delta E_{B}\right)_{g}=\left(\Delta E_{B}\right)_{g}^{E}+\left(\Delta E_{B}\right)_{g}^{M},
\eeq
where
\beq\label{17}
\left(\Delta E_{B}\right)_{g}^{E}=\frac{1}{8 \pi} \sum_{i, j} \sum_{a=1}^{8} \int \frac{d^{3} r_{i} d^{3} r_{j}}{\left|\vec{r}_{i}-\vec{r}_{j}\right|}\left\langle B\left|J_{i}^{0 a}\left(\vec{r}_{i}\right) J_{j}^{0 a}\left(\vec{r}_{j}\right)\right| B\right\rangle,
\eeq
and
\beq\label{18}
\left(\Delta E_{B}\right)_{g}^{M}=-\frac{1}{8 \pi} \sum_{i, j} \sum_{a=1}^{8} \int \frac{d^{3} r_{i} d^{3} r_{j}}{\left|\vec{r}_{i}-\vec{r}_{j}\right|}\left\langle B\left|\vec{J}_{i}^{a}\left(\vec{r}_{i}\right) \cdot \vec{J}_{j}^{a}\left(\vec{r}_{j}\right)\right| B\right\rangle.
\eeq
Here $J_{i}^{\mu a}(x)$ is the color current density of $i$th quark,
\beq\label{19}
J_{i}^{\mu a}(x)=g_{c} \overline{\psi}_{q}(x) \gamma^{\mu} \lambda_{i}^{a} \psi_{q}(x),
\eeq
where $\lambda_{i}^{a}$ are Gell-Mann $SU(3)$ matrices and $\alpha_{c}=g_{c}^{2} / 4 \pi$. Here, we assume that the three quarks in charmed baryons retain the $SU(3)$ symmetry, which is the same case for the strangeness baryons. Then, the color electric contribution and the color magnetic contribution can be given as
\beq\label{20}
\left(\Delta E_{B}\right)_{g}^{E}=\alpha_{c}\left(b_{u u} I_{u u}^{E}+b_{u c} I_{u c}^{E}+b_{c c} I_{c c}^{E}\right),
\eeq
and
\beq\label{21}
\left(\Delta E_{B}\right)_{g}^{M}=\alpha_{c}\left(a_{u u} I_{u u}^{M}+a_{u c} I_{u c}^{M}+a_{c c} I_{c c}^{M}\right).
\eeq

\begin{table}[tb]
\setlength{\tabcolsep}{3mm}
  \caption{The numerical coefficients $a_{i j}$ and $b_{i j}$ are used to calculate the energy contributions                                                                                                                                                                                                                                                                                                                                                                                                                                                                                                                                                                                                                                                 of gluon correction for nucleon and charmed baryon masses.} \label{tab1}	                                                                                                                                                                                                                                                                                                                                                                                                                                                                                                                                                                                                                                                                                                                             \centering                  
   \begin{tabular}{c c c c c c c}
  \hline                                                                                                                                                                                                                                                                                                                                                                                                                       \hline                                                                                                                                                                                                        Baryon &$a_{uu}$&$a_{uc}$&$a_{cc}$&$b_{uu}$&$b_{uc}$&$b_{cc}$\\                                                                                                                                                                                                                                                                                                                                                                                                                                                                                                                                                                                                                                                                    \hline                                                                                                                                                                                                                                                                                                                                                                                                                                                                                                                                                                                                                                            $N$  &-3        &0    &0    &0    &0   &0 \\                                                                                                                                                                                                                                                                                                                                                                                                                                                                                                                                                                                                                                                                                                                                                                                                                                                                                                                                                                                                                                                                                                                                                                                                                                                         $\Lambda_c^+$     &-3        &0 &0 &1 &-2 &1\\                                                                                                                                                                                                                                                                                                                                                                                                                                                                                                                                                                                                                                                                                                                                                                                                                                                                                                                                                                                                                                                                                                                                           $\Sigma_c^+$  &1 &-4 &0 &1 &-2 &1\\                                                                                                                                                                                                                                                                                                                                                                                                                                                                                                                                    $\Xi_{cc}^{++}$ &0 &-4 &1 &1 &-2 &1\\                                                                                                                                                                                                                                                                                                                                                                                                                                                                                                                                       	                                                                                                                                                                                                                                                                                                                                                                                                                                                                                                                                                                                                                                                                                                                                                                                                                                                                                                                                                                                                                                                                                                                                                                                                                                                                                                                                                                                                                                                                                                                                                                                                                                                                                                                                                                                                                                                                                                                                                                                                                                                                                               \hline                                                                                                                                                                                                       \hline
 \end{tabular}
 \end{table}

In Table \ref{tab1}, the coefficients $a_{i j}$ and $b_{i j}$ are shown, {which are related with the expectation values of spin and isospin operators from color current density in Eqs.~(\ref{17}) and (\ref{18}) and are dependent on the species of baryon. They are obtained from the simplified form of Eqs.~(\ref{17}) and (\ref{18}),
\beq\label{17a}
&&\left(\Delta E_{B}\right)_{g}^{E}=\alpha_c\sum_{i,j}\left\langle \sum_{a}\lambda^a_i \lambda^a_j \right\rangle\frac{1}{\sqrt{\pi}R_{i j}}\left[1-\frac{\alpha_{i}+\alpha_{j}}{R_{i j}^{2}}+\frac{3 \alpha_{i} \alpha_{j}}{R_{i j}^{4}}\right]\nn
&&\left(\Delta E_{B}\right)_{g}^{M}=\alpha_c\sum_{i<j}\left\langle \sum_{a}\lambda^a_i \lambda^a_j \sigma_i\cdot\sigma_j\right\rangle \frac{32}{3\sqrt{\pi}R_{i j}^{3}} \frac{1}{\left(3 \epsilon_{i}^{\prime}+m_{i}^{\prime}\right)} \frac{1}{\left(3 \epsilon_{j}^{\prime}+m_{j}^{\prime}\right)}, 
\eeq
and the properties of Gell-Mann $SU(3)$ matrices,
\beq
\left\langle \sum_a(\lambda^a_i)^2 \right\rangle=\frac{16}{3},~~~\left\langle \sum_a\lambda^a_i\lambda^a_j \right\rangle_{i\neq j}=-\frac{8}{3}.
\eeq 
}

Therefore, the quantities $I_{i j}^{E}$ and $I_{i j}^{M}$ are given in the following equations,
\beq\label{22}
I_{i j}^{E}=&&\frac{16}{3 \sqrt{\pi}} \frac{1}{R_{i j}}\left[1-\frac{\alpha_{i}+\alpha_{j}}{R_{i j}^{2}}+\frac{3 \alpha_{i} \alpha_{j}}{R_{i j}^{4}}\right],\nn
I_{i j}^{M}=&&\frac{256}{9 \sqrt{\pi}} \frac{1}{R_{i j}^{3}} \frac{1}{\left(3 \epsilon_{i}^{\prime}+m_{i}^{\prime}\right)} \frac{1}{\left(3 \epsilon_{j}^{\prime}+m_{j}^{\prime}\right)}
,\eeq
with
\beq\label{23}
R_{i j}^{2}=&&3\left[\frac{1}{\left(\epsilon_{i}^{\prime 2}-m_{i}^{\prime 2}\right)}+\frac{1}{\left(\epsilon_{j}^{\prime 2}-m_{j}^{\prime 2}\right)}\right],\nn \alpha_{i}=&&\frac{1}{\left(\epsilon_{i}^{\prime}+m_{i}^{\prime}\right)\left(3 \epsilon_{i}^{\prime}+m_{i}^{\prime}\right)}.\eeq

After all the above energy corrections included, the mass of a charmed baryon in nuclear medium is expressed as:
\beq\label{24}
M_{B}^{*}=E_{B}^{* 0}-\epsilon_{\mathrm{c} . \mathrm{m} .}+\delta M_{B}^{\pi}+\left(\Delta E_{B}\right)_{g}^{E}+\left(\Delta E_{B}\right)_{g}^{M}.
\eeq

Then, the $\Lambda_c^+$ hypernuclei will be studied in QMF model. A single $\Lambda_c^+$ hypernucleus is regarded as a binding system of a $\Lambda_c^+$ baryon and many nucleons which interact via exchanging $\sigma,~\omega$, and $\rho$ mesons. This mechanism of baryon-baryon interaction is originated from the RMF model. Therefore, the Lagrangian of QMF model for  $\Lambda_c^+$ hypernucleus can be written as an analogous form in the RMF model~\cite{shen02,shen06,xing17},
\beq\label{25}
\mathcal{L}_{\mathrm{QMF}}=&&\overline{\psi}_{N}\left[i \gamma^{\mu} \partial_{\mu}-M_{N}^*-g_{\omega N} \omega_{\mu} \gamma^{\mu}-g_{\rho N} \rho_{\alpha \mu} \tau_{\alpha} \gamma^{\mu}-e\frac{(1-\tau_3)}{2} A_{\mu} \gamma^{\mu}\right] \psi_{N}\nn
&&+\overline{\psi}_{\Lambda_c^+}\left[i \gamma^{\mu} \partial_{\mu}-M_{\Lambda_c^+}^*-g_{\omega\Lambda_c^+}
\omega_{\mu} \gamma^{\mu}+\frac{f_{\omega\Lambda_c^+}}{2 M_{\Lambda_c^+}} \sigma^{\mu \nu} \partial_{\nu} \omega_{\mu}-e q_{\Lambda_c^+} A_{\mu} \gamma^{\mu}\right] \psi_{\Lambda_c^+}\nn
&&+\frac{1}{2} \partial_{\mu} \sigma \partial^{\mu} \sigma-\frac{1}{2} m_{\sigma}^{2} \sigma^{2}-\frac{1}{3} g_{2} \sigma^{3}-\frac{1}{4} g_{3} \sigma^{4}\nn
&&-\frac{1}{4} \Omega_{\mu \nu} \Omega^{\mu \nu}+\frac{1}{2} m_{\omega}^{2} \omega_{\mu} \omega^{\mu}+\frac{1}{4} c_{3}\left(\omega_{\mu} \omega^{\mu}\right)^{2}\nn
&&-\frac{1}{4} R_{\alpha \mu \nu} R_{\alpha}^{\mu \nu}+\frac{1}{2} m_{\rho}^{2} \rho_{\alpha \mu} \rho_{\alpha}^{\mu}-\frac{1}{4} F_{\mu \nu} F^{\mu \nu}, \eeq
with
\beq\label{26}
\Omega_{\mu \nu}=&&\partial_{\mu} w_{\nu}-\partial_{\nu} w_{\mu},\nn
R_{\alpha \mu \nu}=&&\partial_{\mu} \rho_{\alpha \nu}-\partial_{\nu} \rho_{\alpha \mu},\nn
F_{\mu \nu}=&&\partial_{\mu} A_{\nu}-\partial_{\nu} A_{\mu}. \eeq
${\psi}_{N}$ and ${\psi}_{\Lambda_c^+}$ are the nucleon and  $\Lambda_c^+$ baryon fields, respectively. $A_\mu$ is the electricmagnetic field for the Coulomb interaction between charged baryons. $M_{N}^{*}$ and $M_{\Lambda_c^+}^{*}$ are the effective masses of nucleon and $\Lambda_c^+$, which can be obtained from the quark potential model. These effective masses are strongly relevant to the magnitudes of $\sigma$ meson in the RMF model. The coupling constants between $\omega, ~\rho$ mesons and nucleons, $g_{\omega N}$ and $g_{\rho N}$, can be determined by the naive quark counting rules, $g_{\omega N}=3g_{\omega q}$ and  $g_{\rho N}=g_{\rho q}$. $g_{\omega q}$ and $g_{\rho q}$ are fixed by the ground-state properties of several doubly magic nuclei. The determination of coupling constants between $\omega$ meson and ${\Lambda_c^+}$ baryon, $g_{\omega\Lambda_c^+}$ and $f_{\omega \Lambda_c^+}$ will be discussed in the next section. $\alpha$ denotes the index of isospin vector. $q_{\Lambda_c^+}$ is the charge of ${\Lambda_c^+}$ baryon with the unit charge $e$. The nonlinear terms of $\sigma$ and $\omega$ mesons are included in this Lagrangian, which can largely improve the descriptions of properties of finite nuclei~\cite{xing16}. In this work, the tensor coupling between $\omega$ meson and ${\Lambda_c^+}$ baryon, $\frac{f_{w\Lambda_c^+}}{2 M_{\Lambda_c^+}} \sigma^{\mu \nu} \partial_{\nu} \omega_{\mu}$ is also introduced following the conventional scheme for $\Lambda$ hypernuclei, where the spin-orbit splittings were very small from the experimental observations~\cite{mares94,shen06,sugahara94}.

In this work, the charmed hypernuclei are regarded as the spherical nuclei and the time-reversal symmetry is assumed. Therefore only time components of the $\omega,~ \rho$, and $A$ fields exist. Furthermore, there is not any contribution from baryon currents. For convenience,  $\omega_0$, $\rho_0,$ and $A_0$ will be replaced by $\omega$, $\rho$, and $A$ in the following. Because of charge conservation, only the third component of the isospin vectors provides a non-vanishing contribution. Here, $\tau_3=-1$ for proton and $\tau_3=1$ for neutron are defined in conventional calculations. With the mean-field approximation, we can get the equations of motion of baryons and mesons by using the Euler-Lagrange equation. The Dirac equations for baryons are given as,
\beq\label{27}
&&\left[i \gamma^{\mu} \partial_{\mu}-M_{N}^{*}-g_{\omega N} \omega \gamma^{0}-g_{\rho N}^{} \rho \tau_{3} \gamma^{0}-e \frac{\left(1-\tau_{3}\right)}{2} A \gamma^{0}\right] \psi_{N}=0,\nn
&&\left[i \gamma^{\mu} \partial_{\mu}-M_{\Lambda_c^+}^{*}-g_{\omega\Lambda_c^+} \omega \gamma^{0}+\frac{f_{\omega\Lambda_c^+}}{2 M_{\Lambda_c^+}} \sigma^{0 i} \partial_{i} \omega-e q_{\Lambda_c^+} A \gamma^{0}\right] \psi_{\Lambda_c^+}=0.
\eeq
The equations of motion for mesons can be obtained by
\beq\label{28}
&&\Delta \sigma-m_{\sigma}^{2} \sigma-g_{2} \sigma^{2}-g_{3} \sigma^{3}=\frac{\partial M_{N}^{*}}{\partial \sigma}\left\langle\overline{\psi}_{N} \psi_{N}\right\rangle+\frac{\partial M_{\Lambda_c^+}^{*}}{\partial \sigma}\left\langle\overline{\psi}_{\Lambda_c^+} \psi_{\Lambda_c^+}\right\rangle,\nn
&&\Delta \omega-m_{\omega}^{2} \omega-c_{3} \omega^{3}=-g_{\omega N}\left\langle\overline{\psi}_{N} \gamma^{0} \psi_{N}\right\rangle- g_{\omega\Lambda_c^+}\left\langle\overline{\psi}_{\Lambda_c^+} \gamma^{0} \psi_{\Lambda_c^+}\right\rangle+\frac{f_{\omega\Lambda_c^+}}{2 m_{\Lambda_c^+}} \partial_{i}\left\langle\overline{\psi}_{\Lambda_c^+} \sigma^{0 i} \psi_{\Lambda_c^+}\right\rangle,\nn
&&\Delta \rho-m_{\rho}^{2} \rho=-g_{\rho N}\left\langle\overline{\psi}_{N} \tau_{3} \gamma^{0} \psi_{N}\right\rangle,\nn
&&\Delta A=-e\left\langle\overline{\psi}_{N} \frac{\left(1-\tau_{3}\right)}{2} \gamma^{0} \psi_{N}\right\rangle- e\left\langle\overline{\psi}_{\Lambda_c^+} q_{\Lambda_c^+} \gamma^{0} \psi_{\Lambda_c^+}\right\rangle.
\eeq
These equations can be solved self-consistently within numerical methods to generate the single-particle energies of baryons and the total energy of charmed hypernucleus.

\section{Result and discussion}
\subsection{Properties of baryons}
The potential parameters $a_q$ and $V_q$ for $u,~d$, and $c$ quarks should be firstly fixed to investigate the properties of baryons. $u$ and $d$ quarks are considered equally due to the very small differences of properties between them, while the $c$ quark is distinguished from them, whose mass is very large. The $u$ or $d$ quark mass in QMF model is adopted from $250-350$ MeV as constituent quark~\cite{xing16,xing17}. Therefore, in order to discuss the influence of quark mass on the properties of baryons, the constituent quark mass for $u$ quark or $d$ quark is taken as $250,~300$, and $350$ MeV, respectively in this work. The corresponding potential parameters $a_u$ and $V_u$ can be derived by fitting the mass and radius of the free nucleon, which have been obtained in our previous work~\cite{xing16,xing17}. For the charm $c$ quark, its mass is chosen as $1300,~1350$, and $1400$ MeV, correspondingly now. The potential parameters $a_c$ and $V_c$ are gained by fitting the experimental masses of $\Lambda_c^+,~\Sigma_c^+$, and $\Xi_{cc}^{++}$ baryons in free space~\cite{patrignani16} with least-squares method.

These parameters are listed in Table \ref{tab2}. For the convenience of latter discussion, the parameters corresponding to $m_u= 250$ MeV in Table \ref{tab2} are named as set A, the parameters corresponding to $m_u= 300~\mathrm{MeV}$ as set B, the parameter corresponding to $m_u= 350~\mathrm{MeV}$ as set C.

\begin{table}[tb]
\setlength{\tabcolsep}{3mm}
  \caption{The potential parameters $a_q$ and $V_q$ for $u$ and $c$ quarks corresponding to $m_u= 250$ MeV as set A, $m_u= 300$ MeV as set B and $m_u= 350$ MeV as set C. } \label{tab2}	
  \centering
\begin{tabular}{c c c c c c c}
	\hline
	\hline
	&$m_u(\rm MeV)$   &$V_u(\rm MeV)$        &$a_u(\rm fm^{-3})$&$m_c(\rm MeV)$     &$V_c(\rm MeV)$      &$a_c(\rm fm^{-3})$\\ \hline
set A&250   &-24.286601&0.579450       &1300          &284.58724          &0.118172                        \\
set B&300   &-62.257187 &0.534296       &1350         &239.53994          &0.117312                          \\
set C&350  &-102.041575&0.495596       &1400          &193.67265          &0.116036                            \\
	\hline
	\hline
\end{tabular}
\end{table}

The masses of charmed baryons, $\Lambda_c^+$, $\Sigma_c^+$, and $\Xi_{cc}^{++}$  in free space generated by set A, set B, and set C are listed to compare with the latest experimental data~\cite{patrignani16} in Table~\ref{tab3}. Meanwhile, the contributions from center-of-mass correction,  pion correction, and gluon correction to the masses of charmed baryons are also shown. It can be found that the charmed baryon masses from the quark potential model almost reproduce their experimental data~\cite{patrignani16} with errors less than $5\%$. Because there are only two degrees of freedom in the confinement potential, $V_c$ and $a_c$. {Furthermore, the masses of $\Xi_{cc}^{++}$ from these three sets reproduce the experimental data better comparing to other two baryons. It is because that two $c$ quarks provide their contributions to $\Xi_{cc}^{++}$, which is more sensitive to the strengths, $a_c$ and $V_c$ in the confinement potentials.}
\begin{table}[tb]
\setlength{\tabcolsep}{3mm}
  \caption{The masses of charmed baryons ($\Lambda_c^+$, $\Sigma_c^+$ and $\Xi_{cc}^{++}$) in free space with set A, set B, and set C parameter sets, compared with the experimental data and various contributions in charmed baryon masses, respectively (the units of all quantities are \rm{MeV}).} \label{tab3}	
  \centering
\begin{tabular}{c c c c c c c c}
	\hline
	\hline
	&Baryon    &$E_B^0$  &$\epsilon_{\rm{cm}}$   &$\delta M_B^\pi$     &$(\Delta E_B)g$      &$M_B^{\rm Theor.}$       &$M_B^{\rm Expt.}$\cite{patrignani16}\\
    \hline
    &$\Lambda_c^+$        &2562.949     &137.904     &-65.172      &-47.747     &2312.126   &2286.46$\pm$0.14  \\
set A&$\Sigma_c^+$         &2562.949     &137.904     &-36.207      &-0.790      &2388.048   &2452.9$\pm$0.4	  \\
    &\multirow{2}*{$\Xi_{cc}^{++}$}      &\multirow{2}*{3737.473}     &\multirow{2}*{96.999}      &\multirow{2}*{-16.293 }     &\multirow{2}*{-15.607}     &\multirow{2}*{3608.574}   &3621.40$\pm$0.72\\
    ~&                     &             &            &             &           &              &$\pm$0.27$\pm$0.14 \\
    &$\Lambda_c^+$        &2558.524     &140.641     &-69.277     &-43.096     &2305.510    &2286.46$\pm$0.14\\
set B&$\Sigma_c^+$         &2558.524     &140.641     &-38.487     &-1.291      &2378.105    &2452.9$\pm$0.4 \\
    &\multirow{2}*{$\Xi_{cc}^{++}$}      &\multirow{2}*{3741.683}     &\multirow{2}*{97.896}      &\multirow{2}*{-17.319}      &\multirow{2}*{-14.588}     &\multirow{2}*{3611.879}    &3621.40$\pm$0.72\\
    ~&                     &             &            &             &           &              &$\pm$0.27$\pm$0.14 \\
    &$\Lambda_c^+$         &2553.749     &141.522     &-72.829     &-39.007     &2300.390    &2286.46$\pm$0.14 \\
set C&$\Sigma_c^+$          &2553.749     &141.522     &-40.461     &-1.674      &2370.092    &2452.9$\pm$0.4\\
    &\multirow{2}*{$\Xi_{cc}^{++}$ }     &\multirow{2}*{3744.384}     &\multirow{2}*{98.099 }      &\multirow{2}*{-18.207}      &\multirow{2}*{-13.640 }     &\multirow{2}*{3614.437}    &3621.40$\pm$0.72\\
    ~&                     &             &            &             &           &              &$\pm$0.27$\pm$0.14\\
    \hline
    \hline
\end{tabular}
\end{table}

\begin{figure}[tb]
	\centering
	\includegraphics[width=11cm]{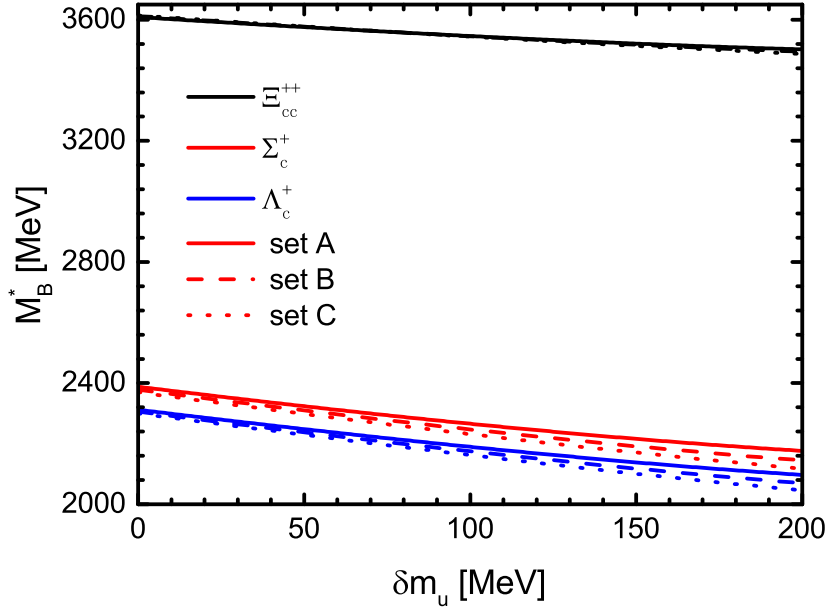}
	\caption{(Color online) The effective masses of charmed baryons, $M_B^*$, for $\Lambda_c^+,~\Sigma_c^+$, and $\Xi_{cc}^{++}$ as  functions of the quark mass corrections $\delta m_u$ with three parameter sets [set A (solid curves), set B (dashed curves), and set C (dotted curves)].}
	\label{fig1}
\end{figure}

The mass of baryon in the nuclear medium $M_B^*$ will vary with nucleon density, because the properties of baryons in the nuclear many-body system are influenced by the surrounding baryons as the famous EMC effect~\cite{aubert83}. In the QMF model, such medium effect is included through the effective quark mass depending on $\sigma$ meson field. In the charmed hypernucleus, the $\sigma$ field only couples with $u$ and $d$ quarks. Therefore, the coupling constant between the $\sigma$ meson and $c$ quark should be taken as zero so that the effective masses of charmed baryons are only affected by $u$ and $d$ quarks. The effective baryon masses are the functions of quark mass corrections $\delta m_u=m_u-m_u^*=g_{\sigma u}\sigma$. In Fig. \ref{fig1}, the effective masses of charmed baryons, $\Lambda_c^+$, $\Sigma_c^+$, and $\Xi_{cc}^{++}$, as functions of $u$ quark mass correction $\delta m_q$ for different parameter sets are plotted. It is found that the effective baryon masses decreased with $\delta m_q$ increasing due to the EMC effect of surrounding baryons. When $\delta m_q$ is zero, the effective masses of these charmed baryons correspond to the free baryon masses. With $\delta m_q$ increasing, the differences of effective masses of $\Lambda_c^+$ and $\Sigma_c^+$ baryons among parameters sets A, B, and C are more obvious than those of $\Xi_{cc}^{++}$ baryon. The reason is that comparing with the $\Xi_{cc}^{++}$ hyperon, there are two light quarks contained in $\Lambda_c^+$ and $\Sigma_c^+$ baryons, which are influenced more by the $\sigma$ meson. It is very similar with the results of $\Lambda,~\Sigma$, and $\Xi$ hyperons in our previous work~\cite{xing17}.

\subsection{Properties of $\Lambda_c^+$ hypernuclei}
The properties of $\Lambda_c^+$ hypernuclei can be studied within QMF model, once the relation between quark mass corrections and effective masses of charmed $\Lambda_c^+$ baryons are derived from quark potential model. The coupling constants between mesons and nucleons have been determined by fitting the ground-state properties of several doubly-magic nuclei in our previous work, {i.e., the binding energies per nucleon and the charge radii of $^{40}\rm{Ca},~^{48}\rm{Ca},~^{90}\rm{Zr}$, and $^{208}\rm{Pb}$ ~\cite{xing16,xing17}. The $\chi^2$ function was defined as 
\beq
\chi^2=\frac{1}{N}\sum_{i=1}^{N}\left(\frac{X^\text{Theo.}_i-X^\text{Exp.}_i}{X^\text{Exp.}_i}\right)^2,
\eeq
with the least square method, where $X$ represents the binding energy, $E/A$ and charge radius, $r_{ch}$ of nuclei. To discuss the mass influences of constituent quark, there were three masses of $u,~d$ quark adopted as $250,~300,~350$ MeV. The corresponding coupling constants between mesons and nucleon were named as QMF-NK1, QMF-NK2, and QMF-NK3, respectively. Their corresponding $\chi^2$ were $3.42\times10^{-5},~2.33\times10^{-5}$, and $1.08\times10^{-5}$. These parameters are listed in Table~\ref{qmfnkp} for the later discussions conveniently.}

\begin{table}[H]
	\centering
	\begin{tabular}{l c c c c c c }
		\hline
		\hline
		                 &~$g_{\sigma}^{q}$~&~$g_{\omega}$~&~$g_{\rho}~$~&~$g_2$~          &~$g_3$ ~     &~$c_3$~     \\
	 	                 &                           &                     &              &$(\rm fm^{-1})$&              &         \\
		\hline		
		QMF-NK1  &5.15871         &11.54726    &3.79601   &-3.52737       &-78.52006  &305.00240 \\
		
		QMF-NK2  &5.09346         &12.30084    &4.04190   &-3.42813       &-57.68387  &249.05654 \\
		
		QMF-NK3  &5.01631         &12.83898    &4.10772   &-3.29969       &-39.87981  &221.68240 \\
		
		\hline
		\hline
	\end{tabular}
	\caption{The coupling constants between mesons and nucleon in QMF-NK1, QMF-NK2, and QMF-NK3 sets.}\label{qmfnkp}
\end{table}

The isospin of $\Lambda_c^+$ baryon is zero, which does not interact with the isovector $\rho$ meson. On the other hand, the coupling strength between $\sigma$ meson and $\Lambda_c^+$ has been included in the effective mass of $\Lambda_c^+$ baryon. Therefore, the $\Lambda_c^+ N$ potential is mainly dependent on the coupling constant between $\omega$ meson and  $\Lambda_c^+$ baryon, $g_{\omega\Lambda_c^+}$ in QMF model. However, there is no specific information about $\Lambda_c^+ N$ interaction at the aspect of experiment. Therefore, we would like to adopt two schemes to fix $g_{\omega\Lambda_c^+}$. The first way is following the method of QMC model and RMF model~\cite{tushima03a,tushima03b,tushima04,tan04,tan04a}, where $g_{\omega\Lambda_c^+}=2/3g_{\omega N}$ according to the naive quark counting rule. In our previous work~\cite{xing16}, the coupling strength between $\omega$ meson and nucleon, $g_{\omega N}$ were taken three values, which were dependent on constituent quark masses. For the convenient of later discussion, the corresponding values of $g_{\omega\Lambda_c^+}$ from naive quark counting rule are called as QMF-NK1C, QMF-NK2C, QMF-NK3C, respectively.

Furthermore, the $\Lambda_c^+ N$ potentials were simulated by Lattice QCD method with different pion masses recently, where the magnitude of $\Lambda_c^+ N$ potential in heavy nuclei, $^{209}_{\Lambda_c^+}\rm {Pb}$,  was just one half of the  $\Lambda N$ potential at the central region by employing the single-folding potential method~\cite{miyamoto18}. {Based on this achievement, we also would like to determine $g_{\omega\Lambda_c^+}$ with the following scheme. Firstly, we make an approximation that the binding energy of $\Lambda_c^+$ in $^{209}_{\Lambda_c^+}\rm Pb$ is one half of that in $^{209}_{\Lambda}\rm Pb$ in QMF model when the Coulomb contribution is turned off. Then the single-$\Lambda$ binding energies at $1s$ state in $^{209}_{\Lambda}\rm Pb$ are calculated within the parameters from our previous work in Ref.~\cite{xing17}. Now the $g_{\omega\Lambda_c^+}$ can be determined through fitting the single-$\Lambda_c^+$ binding energy of $^{209}_{\Lambda_c^+}\rm Pb$.} Finally, three coupling constants between $\omega$ meson and $\Lambda_c^+$ are obtained, which are $g_{\omega \Lambda_c^+}=0.7938g_{\omega N}$ for QMF-NK1C',  $g_{\omega \Lambda_c^+}=0.7806g_{\omega N}$ for QMF-NK2C', and $g_{\omega \Lambda_c^+}=0.7739g_{\omega N}$ for QMF-NK3C'. {In QMF or RMF model, the singe-baryon potential can be written as $U_B=U^B_{S}+U^B_{V}$. The scalar and vector potentials, $U^B_{S}$ and $U^B_{V}$, are related to the scalar meson and vector mesons, respectively. In QMF model, the scalar component has been decided by the quark level. Therefore, when the single-baryon potential is well known, the strength of vector potential is easily obtained. Although the present Lattice QCD simulation only included the contributions from $^1S_0$ and $^3S_1-^3D_1$ channels, they can already represent the basic characters of $ \Lambda_c^+ N$ potentials. It should be a good attempt to connect the density functional theory and lattice calculations with the singe-baryon potential.}  The tensor coupling between $\omega$ meson and $\Lambda_c^+$ baryon will be also included to generate a small spin-orbit splitting in hypernucleus following the conventional way, $f_{\omega\Lambda_c^+}=-g_{\omega \Lambda_c^+}$~\cite{mares94,shen06,sugahara94}. {The detailed values of $g_{\omega \Lambda_c^+}$ from these two schemes are listed in Table~\ref{golc},}
\begin{table}[H]
	\centering
	\begin{tabular}{l c c c c c c }
		\hline
		\hline
		                                      &QMF-NK1C&QMF-NK2C&QMF-NK3C&QMF-NK1C$'$  &QMF-NK2C$'$  &QMF-NK3C$'$     \\
		\hline		
 $g_{\omega \Lambda_c^+}$  &7.69817         &8.20056    &8.55932   &9.16621      &9.60204  &9.93609 \\		
		\hline
		\hline
	\end{tabular}
	\caption{The coupling constants between $\omega$ meson and $\Lambda^+_c$ from the naive quark counting rule and Lattice QCD simulation.}\label{golc}
\end{table}
{It can be found that these coupling constants between $\omega$ meson and $\Lambda^+_c$ baryon are larger than that generated from the $SU(4)$ symmetry in meson-exchange potential~\cite{vidana19}, where $g_{\omega \Lambda_c^+}$ is $5.28191$. It is because that the coupling strengths between the scalar meson and $\Lambda^+_c$ baryon in QMF model are relatively stronger.}

The binding energies per baryon and various radius of single $\Lambda_c^+$ hypernuclei are shown in Table \ref{tab5} within QMF-NK3C and QMF-NK3C' sets from light to heavy mass systems, when the $\Lambda_c^+$ baryon occupies the lowest $1s_{1/2}$ state. The corresponding properties of normal nuclei as the core of the single $\Lambda_c^+$ hypernuclei are also give as comparison. With QMF-NK3C set, the nuclear many-body system becomes more bound when the $\Lambda_c^+$ baryon is included and its charge radius, proton radius, and neutron radius slightly increase. However, the radii of $\Lambda_c^+$ baryon density distribution are smaller than those of proton and neutron in such case. It demonstrates that the $\Lambda_c^+$ baryon is attracted inside the nuclei. These calculations are consistent with the results from RMF model by Tan {\it et al.}~\cite{tan04}. While there are only bound states between $\Lambda_c^+$ baryon and normal nuclei core up to $_{\Lambda_c^+}^{52}\rm {V}$ for single $\Lambda_c^+$ hypernuclei within QMF-NK3C' set, where the coupling constant between $\omega$ meson and $\Lambda_c^+$ baryon is larger than that in QMF-NK3C set. It generates a more repulsive $\Lambda_c^+ N$ potential. Furthermore, Coulomb contributions between $\Lambda_c^+$ baryon and protons are growing with the mass number $A$. Therefore, it can be easily understood that there is no heavy $\Lambda_c^+$ hypernuclei when the $\Lambda_c^+ N$ potential is not so attractive. Actually, this conclusion is very similar with recent work by Miyamoto {\it et al.}, where the $\Lambda_c^+ N$ potential from Lattice simulations was folded to calculate the $\Lambda_c^+$ hypernuclei~\cite{miyamoto18}.  

\begin{table}[tb]
  \centering
\begin{threeparttable}[H]
\setlength{\tabcolsep}{2mm}
  \caption{Binding energies per baryon, $-E/A$, charge radius, $r_{ch}$, and radius (in \rm{fm}) of protons, $r_p$, neutrons, $r_n$ and $\Lambda_c^+$ baryon, $r_{\Lambda_c^+}$, in $\Lambda_c^+(1s_{1/2})$ with QMF-NK3C and QMF-NK3C' sets for $^{16}\rm {O}$, $^{40}\rm {Ca}$, $^{51}\rm {V}$, $^{89}\rm {Y}$, $^{139}\rm {La}$, and $^{208}\rm {Pb}$ and their corresponding single $\Lambda_c^+$  hypernuclei.} \label{tab5}	
  \centering
\begin{tabular}{c c c c c c c c c c c}
	\hline
    \hline
    \multirow{2}{*}{ }&\multicolumn{5}{c}{ QMF-NK3C}&\multicolumn{5}{c}{ QMF-NK3C$'$}\cr
    \cline{2-11}
    &$-E/A$&$r_{ch}$&$r_{p}$&$r_{n}$&$r_{\Lambda_c^+}$&$-E/A$&$r_{ch}$&$r_{p}$&$r_{n}$&$r_{\Lambda_c^+}$\cr\hline
$^{16}\rm {O}$&8.1377&2.7225&2.6042&2.5763&  &8.1377&2.7225&2.6042&2.5763&\cr
$_{\Lambda_c^+}^{17}\rm {O}$&9.1039&2.7298&2.6118&2.5797&1.8199&7.7937&2.7418&2.6244&2.5936&3.1746\cr
$^{40}\rm {Ca}$&8.5916&3.4562&3.3638&3.3141& &8.5916&3.4562&3.3638&3.3141&  \cr
$_{\Lambda_c^+}^{41}\rm {Ca}$&9.0333&3.4630&3.3708&3.3174&2.2599 &8.4159&3.4692&3.3771&3.3252&3.8017\cr
$^{51}\rm {V}$&8.6403& 3.6050&3.5200&3.6127& &8.6403& 3.6050&3.5200&3.6127&\cr
$_{\Lambda_c^+}^{52}\rm {V}$&9.0162&3.6086&3.5237&3.6123&2.3773&8.5047&3.6190&3.5343&3.6246&3.7366\cr
$^{89}\rm {Y}$&8.6990&4.2435&4.1724&4.2923& &8.6990&4.2435&4.1724&4.2923& \cr
$_{\Lambda_c^+}^{90}\rm {Y}$&8.8925&4.2466&4.1755&4.2921&2.9105&&&&& \cr
$^{139}\rm {La}$&8.4276&4.8556&4.7954&4.9826& &8.4276&4.8556&4.7954&4.9826&\cr
$_{\Lambda_c^+}^{140}\rm {La}$&8.5388&4.8565&4.7964&4.9812&3.5325&&&&&\cr
$^{208}\rm {Pb}$&7.8992&5.5037&5.4517&5.6898& &7.8992&5.5037&5.4517&5.6898& \cr
$_{\Lambda_c^+}^{209}\rm {Pb}$&7.9623&5.5052&5.4532&5.6892&4.2618&&&&&\cr
    \hline
    \hline
\end{tabular}
\end{threeparttable}
\end{table}

The energy levels of ${\Lambda_c^+}$ baryons at different angular momenta for various single charmed hypernuclei by using QMF-NK3C and QMF-NK3C' sets are listed in detail in Table \ref{tab6}. The deepest single ${\Lambda_c^+}$ energy level appears in $^{52}_{\Lambda_c^+}\rm {V}$ with parameter set QMF-NK3C at a given angular momentum. It is generated by the competition between the Coulomb repulsion and attractive $\Lambda_c^+ N$ potential.  Both of them become larger for heavy nuclei system. The contribution of $\Lambda_c^+ N$ potential is stronger than that from Coulomb interaction for light hypernuclei, while this situation is opposite at large $A$ case. {This behavior was also shown in the works by Tan {\it et al.}~\cite{tan04a} and Vida\~na {\it et al.}~\cite{vidana19}. The deepest energy levels of ${\Lambda_c^+}$ hypernuclei appeared in $^{41}_{\Lambda_c^+}\rm {Ca}$ from Tan {\it et al.} with RMF model, while in the Model C of Ref.~\cite{vidana19}, the deepest energy level for $1s$ state appeared in $^{91}_{\Lambda_c^+}\rm {Zr}$. Its $\Lambda_c^+ N$ potential was not so attractive among three models.}

It is also found that the spin-orbit splitting of ${\Lambda_c^+}$ hypernuclei is very small. Besides the tensor coupling between $\omega$ and $\Lambda^+_c$, the mass of ${\Lambda_c^+}$ baryons also will influence the spin-orbit force of single ${\Lambda_c^+}$ hypernuclei. When Dirac equation related to ${\Lambda_c^+}$ baryon is reduced to the corresponding Schr\"{o}dinger equation, the spin-orbit force is inversely proportional to the ${\Lambda_c^+}$ baryon mass. Therefore, the spin-orbit force in ${\Lambda_c^+}$ hypernuclei is smaller than that in $\Lambda$ hypernuclei and normal nuclei, {which was consistent with results from RMF model~\cite{tan04a} and perturbative many-body method~\cite{vidana19}}. On the other hand, the $\Lambda_c^+ N$ potential in QMF-NK3C'  is much smaller, where only $1s_{1/2}$ state of ${\Lambda_c^+}$ can exist up to $^{52}_{\Lambda_c^+}\rm {V}$.

\begin{table}[tb]
  \centering
\begin{threeparttable}[H]
\setlength{\tabcolsep}{1.5mm}
  \caption{Energy levels (in MeV) of ${\Lambda_c^+}$ hyperons for $_{\Lambda_c^+}^{17}\rm {O}$, $_{\Lambda_c^+}^{41}\rm {Ca}$, $_{\Lambda_c^+}^{52}\rm {V}$, $_{\Lambda_c^+}^{140}\rm {La}$, and $_{\Lambda_c^+}^{209}\rm {Pb}$ with QMF-NK3C and QMF-NK3C' sets.} \label{tab6}	
  \centering
\begin{tabular}{c c c c c c c c c c c}
	\hline
    \hline
    \multirow{2}{*}{ }&\multicolumn{5}{c}{ QMF-NK3C}&\multicolumn{5}{c}{ QMF-NK3C$'$}\cr
    \cline{2-11}
    &$_{\Lambda_c^+}^{17}\rm {O}$&$_{\Lambda_c^+}^{41}\rm {Ca}$&$_{\Lambda_c^+}^{52}\rm {V}$&$_{\Lambda_c^+}^{140}\rm {La}$ &$_{\Lambda_c^+}^{209}\rm {Pb}$ &$_{\Lambda_c^+}^{17}\rm {O}$&$_{\Lambda_c^+}^{41}\rm {Ca}$&$_{\Lambda_c^+}^{52}\rm {V}$&$_{\Lambda_c^+}^{140}\rm {La}$ &$_{\Lambda_c^+}^{209}\rm {Pb}$ \cr
    \hline
    $1s_{1/2}$  & -24.3013 & -25.8621 & -27.2769 & -21.8919   & -18.0800 &-1.9540&-0.5425&-0.6116& & \\
$1p_{3/2}$  & -16.0223 &-20.4776 & -22.4005&  -19.7552  &  -16.6644 &&&&&\\
$1p_{1/2}$  & -15.9654 & -20.4470&-22.3784&  -19.7470  &  -16.6568 &&&&&\\
$1d_{5/2}$  & -7.4825    &-14.2550 &-16.6196   &  -16.8628  &  -14.5743 &&&&&\\
$1d_{3/2}$  & -7.3925    &-14.1977 & -16.5734  &  -16.8451  &  -14.5595 &&&&&\\
$1f_{7/2}$  &     &-7.5936 & -10.3256  & -13.3801   &  -11.9527 &&&&&\\
$1f_{5/2}$  &     &-7.5100&-10.2519 & -13.3494   &  -11.9285 &&&&&\\
$1g_{9/2}$  &     &-0.7306&-3.7352   &  -9.4215   &  -8.8898 &&&&&\\
$1g_{7/2}$  &     &-0.6267& -3.6360 & -9.3748   &  -8.8543 &&&&&\\
    \hline
    \hline
\end{tabular}
\end{threeparttable}
\end{table}

\begin{figure}[tb]
	\centering
	\includegraphics[width=11cm]{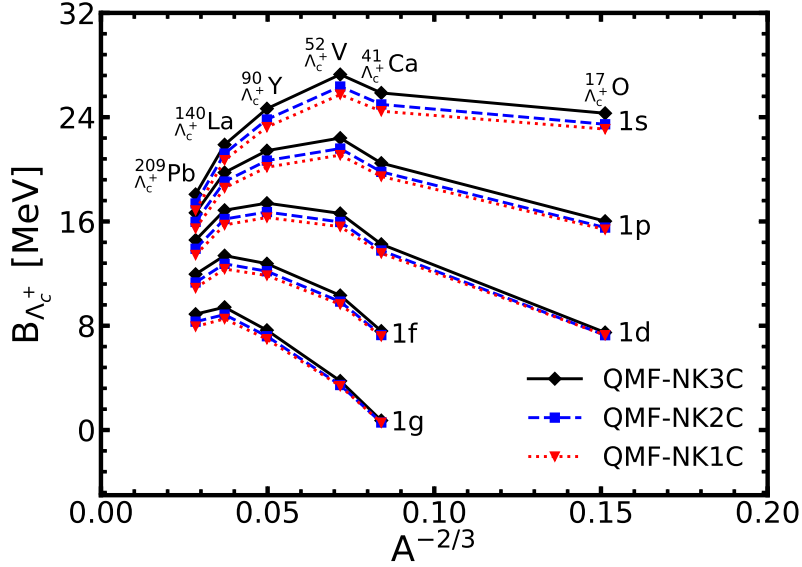}
	\caption{(Color online) The binding energies of single ${\Lambda_c^+}$ hyperons at various angular momenta from $_{\Lambda_c^+}^{17}\rm {O}$ to $_{\Lambda_c^+}^{209}\rm {Pb}$ with three parameter sets [QMF-NK1C (dotted curve), QMF-NK2C (dashed curve), and QMF-NK3C (solid curve)].}
	\label{fig2}
\end{figure}

\begin{figure}[tb]
	\centering
	\includegraphics[width=11cm]{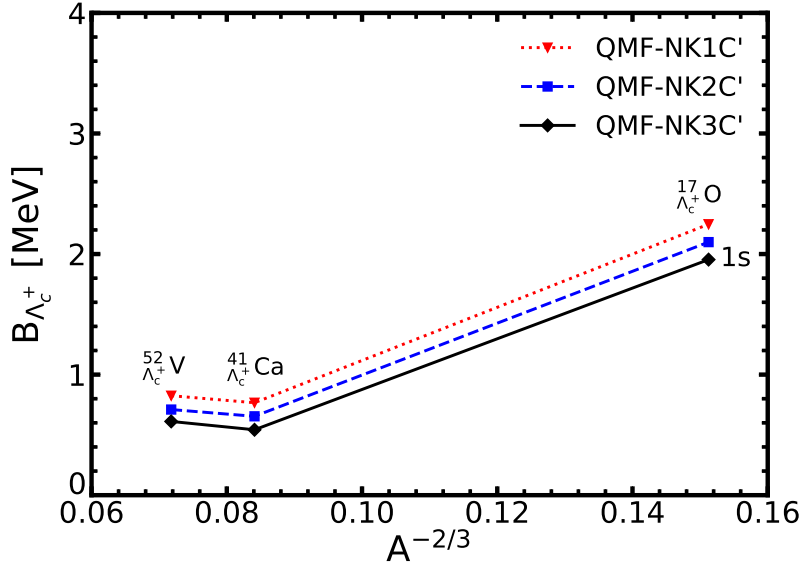}
	\caption{(Color online) The binding energies of single ${\Lambda_c^+}$ hyperons from $_{\Lambda_c^+}^{17}\rm {O}$  to $_{\Lambda_c^+}^{52}\rm {V}$ hypernuclei with three parameter sets [QMF-NK1C' (dotted curve), QMF-NK2C' (dashed curve), and QMF-NK3C' (solid curve)].}
	\label{fig3}
\end{figure}

In Fig. \ref{fig2}, the binding energies of single ${\Lambda_c^+}$ hypernuclei at different angular momenta states are systematically calculated with QMF-NK1C, QMF-NK2C, and QMF-NK3C parameter sets. Their differences among three sets for light and heavy hypernuclei are very small. The differences become obvious at intermediate mass region. The spin-orbit forces of ${\Lambda_c^+}$ hypernuclei are very small now. Therefore, we did not distinguish the spin-orbit partners at a fixed orbital angular momentum here. The corresponding results from QMF-NK1C', QMF-NK2C', and QMF-NK3C' sets are plotted in Fig. \ref{fig3}, where the ${\Lambda_c^+}$ only can occupy the $1s_{1/2}$ state. Furthermore, the binding energies of ${\Lambda_c^+}$ hypernuclei in QMF-NK3C are the largest in the parameter sets which are determined by the naive quark counting rules, while from the lattice simulations, the QMF-NK3C' set generates the smallest binding energies and the differences among the three sets of parameters are almost negligible. It is because that the ${\Lambda_c^+} N$ potentials from lattice simulations are fixed as one half of ${\Lambda} N$ potentials.


\begin{figure}[tb]
	\centering
	\includegraphics[width=11cm]{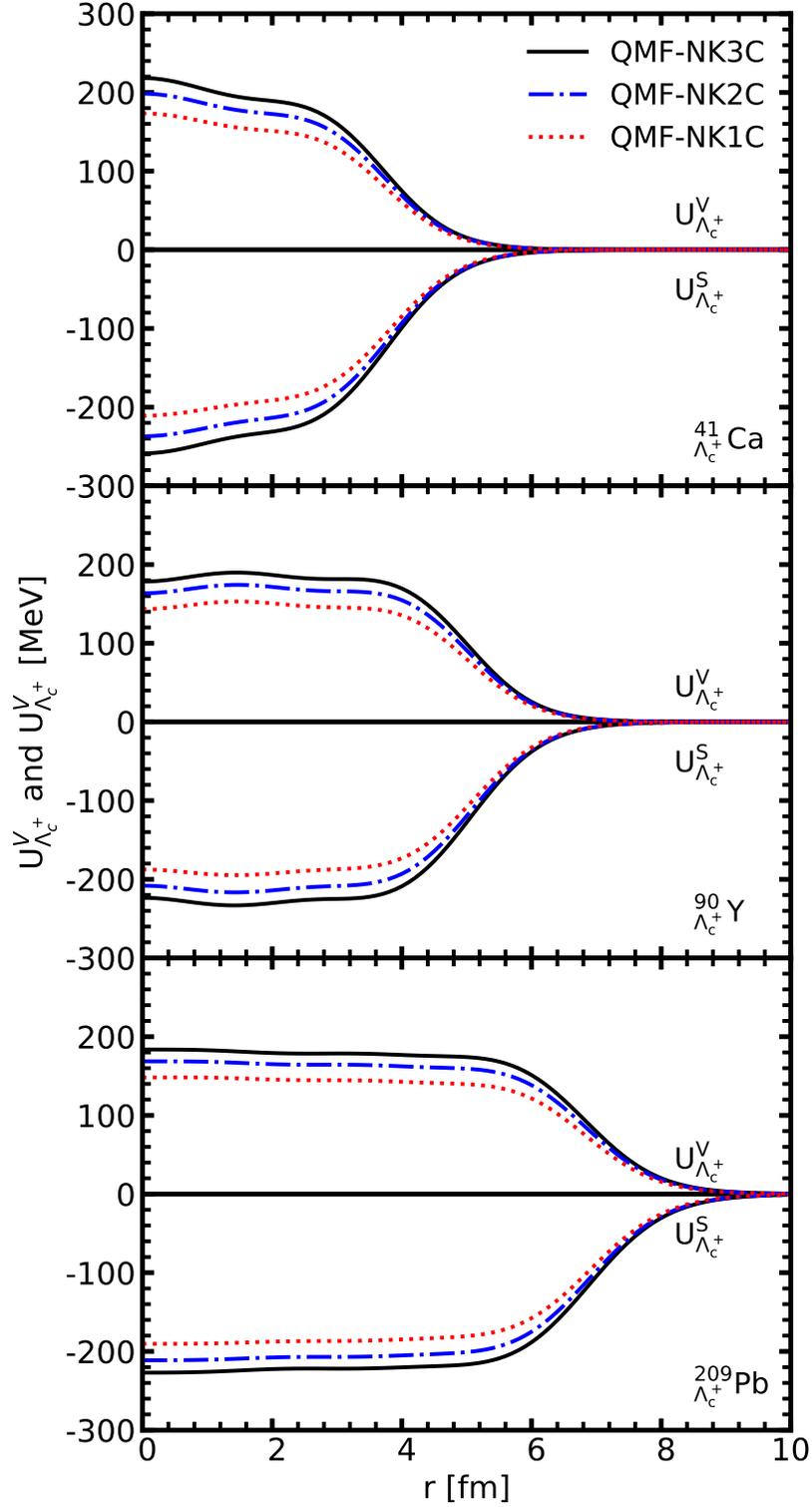}
	\caption{(Color online) The scalar potentials $U_{\Lambda_c^+}^S$ and vector potentials $U_{\Lambda_c^+}^V$ at $\Lambda_c^+$ in $1s_{1/2}$ state for $_{\Lambda_c^+}^{41}\rm {Ca}$, $_{\Lambda_c^+}^{90}\rm {Y}$ and $_{\Lambda_c^+}^{209}\rm {Pb}$ with three parameter sets [QMF-NK1C (dotted curves), QMF-NK2C (dashed-dotted curves), and QMF-NK3C (solid curves)].}
	\label{fig4}
\end{figure}

\begin{figure}[tb]
	\centering
	\includegraphics[width=11cm]{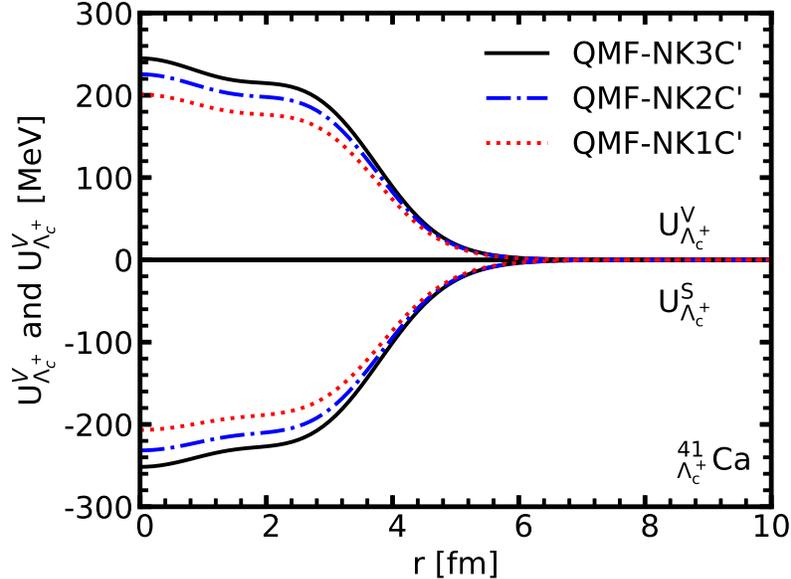}
	\caption{(Color online) The scalar potential $U_{\Lambda_c^+}^S$ and vector potential $U_{\Lambda_c^+}^V$ at $\Lambda_c^+$ in $1s_{1/2}$ state for $_{\Lambda_c^+}^{41}\rm {Ca}$ with three parameter sets [QMF-NK1C' (dotted curves), QMF-NK2C' (dashed-dotted curves), and QMF-NK3C' (solid curves)].}
	\label{fig5}
\end{figure}

The scalar potentials, $U_{\Lambda_c^+}^S$ and vector potentials $U_{\Lambda_c^+}^V$ of $\Lambda_c^+$ baryons at $1s_{1/2}$ states for $_{\Lambda_c^+}^{41}\rm {Ca}$, $_{\Lambda_c^+}^{90}\rm {Y}$, and $_{\Lambda_c^+}^{209}\rm {Pb}$ as functions of their radius are shown in Fig.~\ref{fig4} with QMF-NK1C, QMF-NK2C, and QMF-NK3C sets. These scalar and vector potentials are produced by the $\sigma$ and $\omega$ mesons, respectively. They have the similar magnitudes and lead to total attractive ${\Lambda_c^+ N}$ potentials to bind the ${\Lambda_c^+}$ hypernuclei. This attractive potential at $r=0$ is about $-40$ MeV. The $U_{\Lambda_c^+}^S$ and $U_{\Lambda_c^+}^V$ have the largest magnitude from QMF-NK3C. This is because that the effective $\Lambda_c^+$ mass in set C is the smallest, which can be expressed as $M^*_{\Lambda_c^+}=M_{\Lambda_c^+}+U_{\Lambda_c^+}^S$. The corresponding vector coupling constant, $g_{\omega\Lambda^+_c}$ is the biggest. The ranges of scalar and vector potentials of $\Lambda_c^+$ baryon increase with the mass of $\Lambda_c^+$ hypernuclei. The scalar potential $U_{\Lambda_c^+}^S$ and vector potential $U_{\Lambda_c^+}^V$ from QMF-NK1C', QMF-NK2C', and QMF-NK3C' for $_{\Lambda_c^+}^{41}\rm {Ca}$ are plotted in Fig.~\ref{fig5}. Their behaviors are very similar with the  QMF-NK1C, QMF-NK2C, and QMF-NK3C sets except the smaller vector potentials. In these cases, the $U_{\Lambda_c^+}^S+U_{\Lambda_c^+}^V$ are about $-13$ MeV at central region of charmed hypernuclei, which generated the smaller binding energies.


\begin{figure}[tb]
	\centering
	\includegraphics[width=16cm]{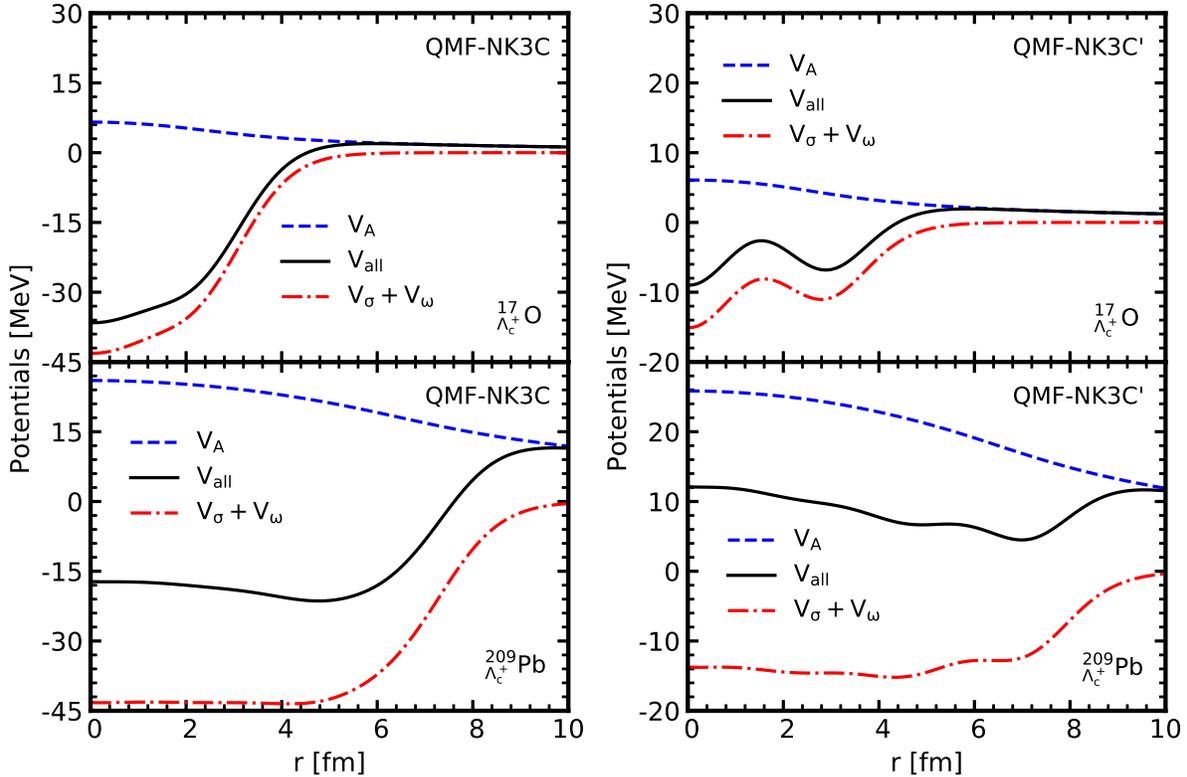}
	\caption{(Color online) The different contributions to the central potential $V_{all}=V_A+V_\sigma+V_\omega$ for $_{\Lambda_c^+}^{17}\rm {O}$ and $_{\Lambda_c^+}^{209}\rm {Pb}$ within QMF-NK3C and QMF-NK3C'. The dashed lines are from the Coulomb contribution, the dashed-dotted curves represent the sum of contributions from $\sigma$ meson and $\omega$ meson, and the total ones are given as solid curves.}
	\label{fig6}
\end{figure}

Actually, the properties of ${\Lambda_c^+}$ baryons in ${\Lambda_c^+}$ hypernuclei are determined by the total potentials from $\sigma$ meson, $\omega$ meson, and Coulomb field. In Fig. \ref{fig6}, the contributions to ${\Lambda_c^+} N$ potential from $\sigma$ and $\omega$, $V_\sigma+V_\omega$, the Coulomb interaction, $V_A$, and the total, $V_{all}=V_\sigma+V_\omega+V_A$ are shown for  $_{\Lambda_c^+}^{17}\rm {O}$ and $_{\Lambda_c^+}^{209}\rm {Pb}$ within QMF-NK3C (left panel) and QMF-NK3C' set (right panel). It can be found that the sums of $\sigma$ and $\omega$ potentials for $_{\Lambda_c^+}^{17}\rm {O}$ and $_{\Lambda_c^+}^{209}\rm {Pb}$ both are around $-45$ MeV by using the QMF-NK3C set. However, the contributions provided by Coulomb force in these two hypernuclei are completely different, which are around $7$ MeV and $26$ MeV for $_{\Lambda_c^+}^{17}\rm {O}$ and $_{\Lambda_c^+}^{209}\rm {Pb}$, respectively. Therefore, the total potential of $_{\Lambda_c^+}^{209}\rm {Pb}$ is much smaller than that of  $_{\Lambda_c^+}^{17}\rm {O}$, which generates the deeper single ${\Lambda_c^+}$ energies for light charmed hypernuclei. In QMF-NK3C' set, there are also the similar behaviors. Now, the $V_\sigma+V_\omega$ is just about $-15$ MeV. In this case, the strong repulsion from the Coulomb interaction cannot generate any bound state for heavy ${\Lambda_c^+}$ hypernuclei.

\section{conclusion}
The single $\Lambda^+_c$ hypernuclei were studied within the quark mean-field (QMF) model. Firstly, a baryon was regarded as a combination composed by three constituent quarks, which were confined by central harmonics oscillator potentials with Dirac vector-scalar mixing form. Furthermore, the pion and gluon corrections were also included to treat the baryons from strong interaction more realistically.
The strengths of the confinement potentials for $u,~d,~c$ quarks, were fixed by the masses and radii of baryons from the observations after considering three different constituent quark masses.

At the aspect of nuclear many-body system, the baryons interacts with each other in the hypernucleus via exchanging the scalar and vector mesons between the quarks in different baryons. The coupling constants between the vector mesons and $u, ~d$ quarks have been obtained by fitting the ground-state properties of several double magic nuclei. The $\Lambda^+_c N$ potential was very significant to study the properties of  single $\Lambda^+_c$ hypernuclei, which were decided by the coupling strength between $\omega$ meson and $\Lambda^+_c$ baryon. Therefore, two schemes were adopted in this work. The first one was that the naive quark counting rule was adopted, where $g_{\omega \Lambda^+_c}=2/3g_{\omega N}$. In the second way, the conclusion of latest lattice simulations provided a good reference, {which pointed out that the $\Lambda^+_c N$ potential was just one half of $\Lambda N$ potential in $_{\Lambda_c^+}^{209}\rm {Pb}$ with single-folded potential method}. Finally, two kinds of parameter sets were obtained, named as QMF-NK1C, QMF-NK2C, QMF-NK3C, and QMF-NK1C', QMF-NK2C', QMF-NK3C', respectively with different constituent quark masses.

The properties of single $\Lambda^+_c$ hypernuclei were systematically calculated from light to heavy mass region. The nuclear many-body systems became more bound when the $\Lambda^+_c$ baryon were included for QMF-NK1C, QMF-NK2C, and QMF-NK3C parameter sets. The rms radii of $\Lambda^+_c$ baryon density distribution were much smaller than those of protons and neutrons. It means that the $\Lambda^+_c$ baryon was inside of the $\Lambda^+_c$ hypernuclei. When the lattice simulation results were used, the $\Lambda^+_c N$ potential did not bind so deeply. There was not bound state of heavy $\Lambda^+_c$ hypernuclei due to the strong repulsive contribution from Coulomb force up to $A\sim 50$. These results were consistent with the recent calculations by {RMF model, HAL QCD group, and perturbative many-body method.}

The single $\Lambda^+_c$ energies were also studied when the $\Lambda^+_c$ baryons were fixed at particular angular momenta. The $\Lambda^+_c$ baryon can occupy very high angular momentum state when the coupling constants between $\omega$ meson and  $\Lambda^+_c$ baryon were adopted by naive quark counting rules.  Meanwhile, there were only $1s_{1/2}$ states with QMF-NK1C', QMF-NK2C', and QMF-NK3C' sets, where shallow $\Lambda^+_c N$ potentials were generated by scalar, vector mesons, and Coulomb field from HAL QCD data.

The strength of $\Lambda^+_c N$ potential is the significant quantity in investigating the properties of single $\Lambda^+_c$ hypernuclei, which cannot be determined by experimental observations very well now. In this work, two schemes were adopted, which have very large differences for heavy nuclei system. The relevant experiments about $\Lambda^+_c$ hypernuclei  are expected to be done, especially in the heavy mass region to determine the magnitude of $\Lambda^+_c N$ potential.

\section*{Acknowledgments}
J. Hu is very grateful to Ying Zhang for a careful reading of this manuscript. This work was supported in part by the National Natural Science Foundation of China (Grant No. 11775119 and No. 11675083) and the Natural Science Foundation of Tianjin.

 \end{document}